\documentclass[showpacs,preprint]{revtex4}

\usepackage{amsmath,amssymb}
\usepackage{graphicx}
\usepackage{dcolumn}
\usepackage{bm}

\begin{document}

\newcommand{\be}{\begin{equation}}
\newcommand{\ee}{\end{equation}}
\newcommand{\bea}{\begin{eqnarray}}
\newcommand{\eea}{\end{eqnarray}}
\newcommand{\mc}{\mathcal}
\newcommand{\eps}{\varepsilon}
\newcommand{\s}{\sigma}
\renewcommand{\d}{\text {d}}
\newcommand{\e}{\text {e}}
\newcommand{\ds}{\Delta\sigma}
\newcommand{\hw}{h^{\rm W}}
\newcommand{\sw}{\sigma^{\rm W}}
\newcommand{\<}{\langle}
\renewcommand{\>}{\rangle}
\newcommand{\de}{\partial}
\newcommand{\eq}{\text{ eq}}
\newcommand{\sign}{\text{ sign}}
\newcommand{\ua}{\uparrow}
\newcommand{\da}{\downarrow}

\title{Nonequilibrium critical dynamics of the ferromagnetic Ising
model with Kawasaki dynamics}

\author{Claude Godr\`eche}
\affiliation{Service de Physique de l'\'Etat Condens\'e, CEA Saclay,
91191 Gif sur Yvette cedex, France}

\author{Florent Krz\c{a}ka{\l}a and Federico Ricci-Tersenghi}
\affiliation{Dipartimento di Fisica, INFM (UdR Roma I and SMC center),
Universit\`a di Roma ``La Sapienza'', P.~A.~Moro 2, 00185 Roma, Italy}

\date{January 19, 2004}

\begin{abstract}
We investigate the temporal evolution of a ferromagnetic system of
Ising spins evolving under Kawasaki dynamics from a random initial
condition, in spatial dimensions one and two. We examine in detail the
asymptotic behaviour of the two-time correlation and response
functions. The linear response is measured without applying a field,
using a recently proposed algorithm. For the chain at vanishingly
small temperature, we introduce an accelerated dynamics which has the
virtue of projecting the system into the asymptotic scaling
regime. This allows us to revisit critically previous works on the
behaviour at large time of the two-time autocorrelation and response
functions.  We also analyse the case of the two-dimensional system at
criticality.  A comparison with Glauber dynamics is performed in both
dimensionalities, in order to underline the similarities and
differences in the phenomenology of the two dynamics.
\end{abstract}


\maketitle

\section{Introduction}

The kinetics of ferromagnetic spin systems evolving after an initial
quench from a high temperature disordered initial condition to a final
temperature, equal or below the critical temperature, is a well
investigated field (see e.g.\ ref.~\cite{bray} for a review).
However, only recently has the emphasis been put on the dynamics of
two-time quantities, such as the correlation and response functions,
or the fluctuation-dissipation violation ratio, with the aim of
quantifying the distance of the system to equilibrium during its
temporal evolution~\cite{janssen,huse1,cuku1,cuku2,barrat,berthier,
gl1D,lip,gl2D,malte,cala,calb,pico,mayer1,mayer2,cris,sastre} (see
e.g.\ \cite{gl02} for a brief review). Most of these studies focus on
the non-conserved order parameter case, where at each time step a
single spin is updated according to the rules of Glauber dynamics.  An
overall coherent picture of this field has by now emerged, though some
controversies remain~\cite{salerno2,pleimling,malte2,salerno3}.

Much fewer studies have been devoted so far to the same questions,
namely the temporal evolution of two-time quantities, for the case of
conserved order parameter dynamics. Restricting to a system of
discrete spins, the rules of Kawasaki dynamics~\cite{kawa} now consist
in choosing two adjacent opposite spins, and exchange them with a rate
depending on the energy difference between the initial and final
configurations. In the recent past the question of the long-time
behaviour of the autocorrelation function for conserved dynamics has
already been addressed~\cite{maj,alexander,yeung}. In particular
predictions have been given for the values of the autocorrelation
exponents $\lambda$ and $\lambda_c$ governing the decay of the
autocorrelation function at large temporal separations, respectively
in the low temperature phase, and at criticality. Finally, a very
recent work addresses the question of the response for the
Ising-Kawasaki chain in the low temperature scaling
regime~\cite{salerno1}. The prediction made in this reference states
that the fluctuation-dissipation plot (that is the relationship
between integrated response and correlation) of this case is identical
to that of the Glauber non-conserved case, which is itself known
analytically~\cite{gl1D,lip}. This result is rather surprising because
it would imply some kind of ``super universal" behaviour in the
relationship between correlation and response in the asymptotic
regime.

The aim of the present work is to revisit and extend these former
studies. We investigate the behaviour of the two-time correlation and
response functions for an Ising spin system, both in one and two
dimensions, with Hamiltonian
\be
{\cal H}=-\sum_{\<i,j\>}\sigma_i\sigma_j
\label{hamilt}
\ee
where $\<i,j\>$ are nearest neighbours, evolving under Kawasaki
dynamics after a quench of the system from a disordered high
temperature initial condition to the critical temperature.

We first give the method used in this paper in order to compute the
linear response. This method, due to Chatelain~\cite{chatelain}, and
later on clarified in~\cite{fede}, will be used both for the
one-dimensional and two-dimensional cases. We then describe the rules
of an accelerated dynamics for the Ising-Kawasaki chain corresponding
to the formal limit $T\to0$, which is nevertheless faithful, i.e.\
reproduces exactly the results that would be obtained with the usual
rules of Kawasaki dynamics for vanishingly small temperatures.  We
finally present the results of extensive numerical computations of the
autocorrelation and response functions, first for the case of a chain,
then for the two-dimensional system at criticality.

\section{Measuring the linear response without applying a field in
 Kawasaki dynamics}

In this section we give an analytical expression of the response
function to an infinitesimal field, for a ferromagnetic system
evolving under Kawasaki dynamics, and show how this quantity can be
measured.  We follow the lines of reasoning of
refs.~\cite{chatelain,fede}, which are devoted to the same question
for single-spin flip dynamics.

\subsection{Kawasaki dynamics with heat bath rule}

Hereafter, time $t$ is discrete and counts the number of spin exchange
attempts, and not the number of Monte Carlo sweeps. In order to define
a response function, an external perturbing field $h_i$ is applied on
any site $i$, and, in presence of the field, the Hamiltonian is
changed to ${\cal H}-\sum h_i\s_i$.

Kawasaki rules consist in updating a pair of two opposite adjacent
spins $\s_i=-\s_j$ (we will always omit the $\delta_{\s_i,-\s_j}$
factor in the following), with heat bath probabilities
\be
{\cal P}(\s_i=\s,\s_j=-\s) = \frac{\exp[\beta \s (\hw_{ij}+h_i-h_j)]}
{2 \cosh[\beta (\hw_{ij}+h_i-h_j)]}\;,
\ee
where $\beta$ is the inverse temperature, and the Weiss field
$\hw_{ij}$ takes into account the effect of neighbours on the couple
of spins which are updated. For a generic 2-spins interaction
Hamiltonian we have
\be
\hw_{ij} = \sum_{k \in \partial i \backslash j} J_{ik}\,\s_k -
\sum_{l \in \partial j \backslash i} J_{jl}\,\s_l \;,
\ee
where $\de i\backslash j$ represents the set of neighbours of $i$,
with $j$ excluded. For example, in one dimension, with
Hamiltonian~(\ref{hamilt}), we have $\hw_{i,i+1} =\s_{i-1}-\s_{i+2}$.

\subsection{Response function}

Following strictly the notation of ref.~\cite{fede}, we consider
systems made of $N$ Ising spins, where the autocorrelation and the
response functions are defined as
\be
C(t,s) = \frac1N \sum_{i=1}^N \< \s_i(t) \s_i(s) \>\;, \qquad
R(t,s) = \frac1N \sum_{i=1}^N \frac{\de\<\s_i(t)\>}{\de h_i(s)}\;,
\ee
with $\<\,\cdot\,\>$ representing the average over thermal histories.

We concentrate on the integrated response function, or susceptibility
\be
\chi(t,s) = T \int_{s}^t \d u\,R(t,u) \quad ,
\label{zfc}
\ee
where the temperature $T$ has been added in the usual definition in
order to simplify the notation and to have a well defined expression
in the $T\to 0$ limit.

Denoting by $I(t)$ and $J(t)$ the indices of the two spins to be
updated at time $t$, the expectation value of the $k$-th spin at time
$t$ is given by
\be
\<\s_k(t)\> = {\rm Tr}_{\vec\s(t')}\left[ \s_k(t) \prod_{t'=1}^t
 W_{I(t')J(t')}\Big( \vec\s(t') | \vec\s(t'\!-\!1) \Big) \right]\;,
\label{eq1}
\ee
where $\vec\s$ is the vector of the $N$ spins configuration, the trace
is over all the histories $\vec\s(t')$ with $1 \le t' \le t$, and the
transition probability are given by
\be
W_{ij}(\vec\s | \vec\tau) = \frac{\exp[\beta \s_i (\hw_{ij}+ h_i-h_j)]}
{2 \cosh[\beta (\hw_{ij}+h_i-h_j)]} \prod_{k \neq i,j}
\delta_{\s_k,\tau_k}\;.
\ee
Note that $\hw_{ij}(\vec\s) = \hw_{ij}(\vec\tau)$, because the Weiss
field does not depend on the value of spins at sites $i$ and $j$.
Since the transition probability $W_{ij}$ only depends on the
perturbing fields on sites $i$ and $j$, one has
\be
\left. \frac{\de W_{ij}(\vec\s | \vec\tau)}{\de h_k} \right|_{h=0} =
\beta W_{ij}(\vec\s | \vec\tau) \Big[ \delta_{i,k} (\s_i - \sw_{ij})
+ \delta_{j,k} (\s_j + \sw_{ij}) \Big] \;,
\ee
where we have defined $\sw_{ij} \equiv \tanh(\beta \hw_{ij})$.

Now, if on site $k$ an infinitesimal probing field $h_k$ is switched
on at time $s$ (i.e.\ $h_k(t) = h\,\theta(t-s)$), all transition
probabilities with index $k$ (and only these ones) will depend on the
perturbing field for times larger than $s$. Then differentiation of
eq.~(\ref{eq1}) with respect to this field yields the integrated
response
\begin{multline}
\chi_{lk}(t,s) = T \left.\frac{\partial \<\s_l(t)\>}{\partial h_k}
\right|_{h=0} = {\rm Tr}_{\vec\s(t')} \left[ \s_l(t) \prod_{t'=1}^t
W_{I(t'),J(t')} \Big(\vec\s(t') | \vec\s(t'\!-\!1)\Big) \right.\\
\left.\sum_{u=s+1}^t \left[
\delta_{I(u),k}\,\left(\s_k(u) - \sw_{I(u)J(u)}(u)\right) +
\delta_{J(u),k}\,\left(\s_k(u) + \sw_{I(u)J(u)}(u)\right)
\right]
\right]
\end{multline}
which can be simply written as a correlation function
\be
\chi_{lk}(t,s) = \<\s_l(t)\,\ds_k(t,s)\>\;,
\label{resp}
\ee
where
\be
\ds_k(t,s) = \sum_{u=s+1}^t
\delta_{I(u),k} \Big[\s_k(u) - \sw_{I(u)J(u)}(u) \Big] +
\delta_{J(u),k} \Big[\s_k(u) + \sw_{I(u)J(u)}(u) \Big]\;.
\ee
%

\subsection{Computing the linear response in a simulation}
\label{computing_resp}

Let us note that calculating the linear response in a numerical
simulation with no perturbing field using eq.~(\ref{resp}) is as easy
as measuring a correlation function. One has to keep track of the
vector $\ds_k(t,s)$ that should be updated for all times between $s$
and $t$ (i.e.\ when the ``ghost" field is switched on). At each of
these times, one has to:
\begin{itemize}
\item compute $\hw_{ij}$ and $\sw_{ij} \equiv \tanh(\beta \hw_{ij})$;
\item update spins $\s_i$ and $\s_j$ according to the heat bath
probability (with no external field)
\[
{\cal P}(\s_i=\s,\s_j=-\s) = \frac{\exp(\beta \s
 \hw_{ij})}{2\cosh(\beta \s \hw_{ij})}\;;
\]
\item increment $\ds_i$, $\ds_j$ as
\bea
\ds_i &\to& \ds_i + \s_i - \sw_{ij} \quad , \label{update_ds1} \\
\ds_j &\to& \ds_j + \s_j + \sw_{ij} \quad . \label{update_ds2}
\eea
\end{itemize}
It is important to note at this point that contributions to the
increment $\ds_i$ come either when the updated spin flips,
or {\it when it does not}, keeping its previous value.
This is the reason why one can speak of a ``ghost" field
in this method (see above).
An illustration of this fact is encountered in 
section~\ref{accel} B below.

The use of eq.~(\ref{resp}) allows one to compute, in a single run,
the integrated response~(\ref{zfc}) (zero-field cooled magnetization),
as well as the thermoremanent magnetization,
\be
\rho(t,s) = T \int_{0}^s \d u\,R(t,u) ,
\label{trm}
\ee
both for many different values of $s$.

\section{Accelerated dynamics of the Ising-Kawasaki chain in
the zero temperature limit}
\label{accel}

It is well known that Kawasaki dynamics at zero temperature rapidly
brings the one-dimensional system to a blocked state, where the
distance between any couple of domain walls is at least
2~\cite{cks,smedt}. In such a situation any spin-exchange move would
cost an energy $\Delta E = 4$, and is therefore forbidden. For example
$\ua\ua\ua\ua\da\da\da\da\;\to\;\ua\ua\ua\da\ua\da\da\da$ would create
two domain walls. Such a process is called an 
{\em evaporation}~\cite{cks}.

However, at an infinitesimal temperature $T=\eps$, the evolution may
eventually undergo such an evaporation with probability
$1/(1+\e^{4/\eps})$, hence on a time scale $O(\e^{4/\eps})$, which
diverges for $\eps\to 0$. After each evaporation process the dynamics
proceeds by {\em diffusion}, that is by moves with energy cost $\Delta
E = 0$ (e.g.\
$\da\da\da\ua\da\da\da\da\;\to\;\da\da\da\da\ua\da\da\da$), until a
{\em condensation} process occurs, corresponding to a move with energy
cost $\Delta E = -4$, which brings the system back to a new blocked
state (e.g.\
$\da\da\da\ua\da\ua\ua\ua\;\to\;\da\da\da\da\ua\ua\ua\ua$). The
diffusion and condensation processes take place on scales of time
$O(1)$, i.e.\ much smaller than the typical time between two
consecutive evaporations, and therefore can be considered as
instantaneous if time is counted in units of $\tau=\e^{4/\eps}$, when
$\eps\to 0$.

We exploit the strong separation of time scales between evaporation
and diffusion/con\-densation processes, and we simulate the dynamics
of the Ising-Kawasaki chain in the limit $T=\eps\to0$ with the
following accelerated dynamics. Time in this accelerated dynamics is
counted in units of $\tau=\e^{4/\eps}$.

\subsection{Rules for the accelerated dynamics}

As long as the system is not in a blocked state, we use the usual
$T=0$ Kawasaki dynamics where only spin exchanges with $\Delta E \le
0$ are accepted. During these lengths of time, since the only
processes occurring, diffusion and condensation, are actually
instantaneous on the scale $\tau=\e^{4/\eps}$, time is not increased
at all.

When the system is in a blocked state (with a number $n$ of domain
walls), we choose randomly a domain wall, we exchange the two spins on
the sides of the domain wall (evaporation), and we increase the time
by $\Delta t = 1/n$.

The choice for $\Delta t$ can be understood as follows. In a Monte
Carlo simulation at $T=\eps$, the probability of accepting an
evaporation process being approximately equal to $\e^{-4/\eps}$, on
average, a number $O(\e^{4/\eps})$ of tries will be necessary before a
success. In each Monte Carlo sweep (MCS) $n$ tries are made, since the
only couples of spins which satisfy the requirement $\s_i=-\s_j$ are
those around a domain wall. Thus the typical number of MCS done before
an evaporation process takes place is $\tau/n$, i.e.\ $\Delta t = 1/n$
in units of $\tau$.

\subsection{Rules for updating $\ds$ during the accelerated dynamics}

Being a quantity integrated over time, $\ds_k(t,s)$ gets contributions
from both fast and slow processes.

Moreover, for $T\!\to\!0$, eqs.~({\ref{update_ds1}, \ref{update_ds2})
can be simplified to $\ds_i \to \ds_i +\{2\s_i,\s_i,0\}$, and $\ds_j
\to \ds_j +\{2\s_j,\s_j,0\}$, for evaporation, diffusion and
condensation processes, respectively.

Between two evaporation processes, there is however an additional
contribution to take into account (see eq.~(\ref{add}) below), which
is not apparent in eqs.~(\ref{update_ds1}, \ref{update_ds2}).  This is
a direct consequence of the fact, mentioned above, that $\ds_k(t,s)$
gets contributions of updated spins, even when they do not flip.
Consider a blocked configuration, where all the spins are aligned with
their local fields, $\s_i = \sign(\hw_{ij})$ and $\s_j =
-\sign(\hw_{ij})$.  Since the time spent in each blocked state becomes
infinite for $\eps \to 0$, a non trivial limit for the integrated
quantity $\ds$ is generated.  Indeed, working at temperature $T=\eps$,
and, noting that in a blocked state $\hw_{ij} = 2\s_i$, we have,
\be
\sw_{ij} = \tanh(\hw_{ij}/\eps) \approx \sign(\hw_{ij}) \left[ 1
- 2 \e^{-2 |\hw_{ij}|/\eps} \right]
=\sigma_i\left (1-\frac{2}{\tau}\right ),
\ee
hence
\be
\s_i - \sw_{ij} \approx \frac{2 \s_i}{\tau}.
\ee
Then the summation over $\tau/n$ MCS of the last quantity gives a
finite limit when $\eps \to 0$ and $\tau \to \infty$. In practice this
means that, just before leaving a blocked state by an evaporation
process, all the $\ds_i$ for spins close to a domain wall must be
updated with the following rule
\bea
\ds_i \to \ds_i + \frac{2 \s_i}{n}.
\label{add}
\eea

Some final comments on the accelerated dynamics described in this
section are in order.
\begin{itemize}
\item
This dynamics is faithful, i.e.\ it reproduces {\em exactly} the
results that would be obtained by a standard Monte Carlo simulation at
finite temperature $T=\eps$, in the limit $\eps\to0$.  This
equivalence will be illustrated below on the example of the mean
length of domains.  It is also very efficient, since it requires much
less computational efforts than the standard Monte Carlo dynamics.
\item
Its definition can be extended to any spatial dimension.
\item
Finally, one may wonder how this dynamics compares to the effective
dynamics of refs.~\cite{cks,cordery}, which is only defined in one
dimension.  The spirit of the later is to trace upon all events
occurring between the instant of time a spin detaches from a domain
and that when it reaches the neighbouring domain.  The accelerated
dynamics introduced in this section does not trace upon these events.
However doing so --in one dimension only-- would lead to the dynamics
of~\cite{cks,cordery}, and allow a faster computation of the average
domain length $L(t)$, and with little more work, of the
autocorrelation.  In contrast, tracing upon these events is much more
subtle for the computation of the response, and would deserve further
study.  Otherwise stated, it is not clear to us for the time being
whether the method of~\cite{cks,cordery} can be used for the
computation of the response.  The difficulty comes from the fact that
$\ds$ gets contributions even when spins are updated without changing
their value.
\end{itemize}

\section{Results for the dynamics of the Ising-Kawasaki
chain in the zero-temperature limit}

In this section we report the results of extensive numerical
simulations, using the methods described in the previous sections. We
are interested in the behaviour of observables in the low-temperature
scaling regime defined by $1\ll t\ll t_{\rm eq}$, where the
equilibration time $t_{\rm eq}$ is related, at inverse temperature
$\beta=1/\eps$, to the equilibrium correlation length $\xi\approx
\e^{2\beta}/2$ by $t_{\rm eq}\sim\xi^5$~\cite{glKawa}.  This regime is
naturally attained by the accelerated dynamics.

\subsection{Mean domain length}

\begin{figure}[htb]
\begin{center}
\includegraphics[width=.75\linewidth]{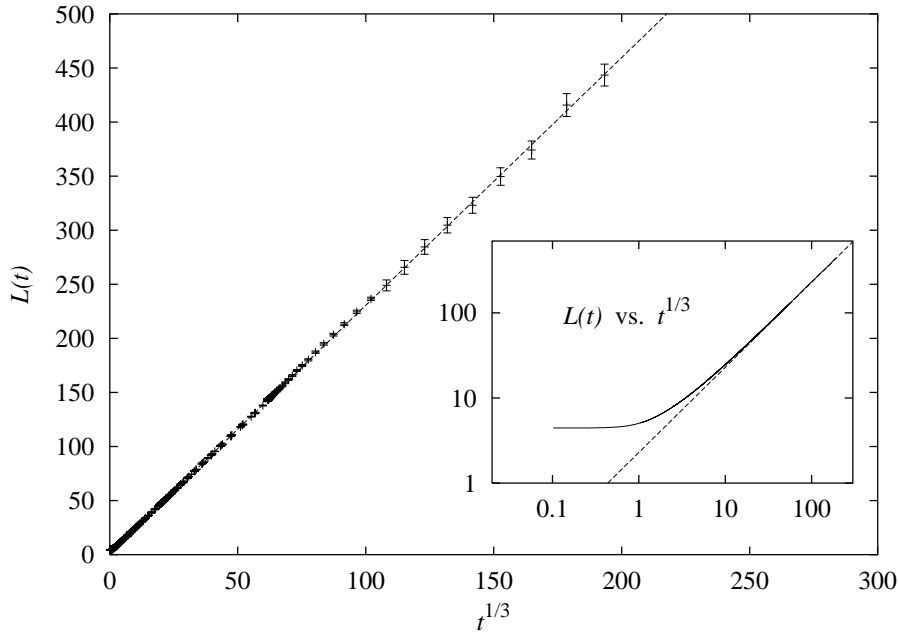}
\caption{Mean domain length $L(t)$ against $t^{1/3}$ for the
Ising-Kawasaki chain obtained by the accelerated dynamics for
vanishingly small temperature. Inset: same data on a log-log scale.}
\label{f1}
\end{center}
\end{figure}

The mean domain length is related to the energy $E(t)$ of the chain at
time $t$ by
\be
L(t)=\frac{2}{1+E(t)}.
\ee
It is well known that, in the low-temperature scaling regime, $L(t)$
scales as
\be
L(t)\sim\left(\frac{t}{\tau}\right)^{1/3},
\label{Lt}
\ee
where $\tau=\e^{4\beta}\sim\xi^2$. This scaling is illustrated in
ref.~\cite{glKawa}, where for increasing values of $\tau$ (i.e.\
decreasing temperatures) a linear master curve is found when $L(t)$ is
plotted against $(t/\tau)^{1/3}$.

We measured $L(t)$ using the accelerated dynamics of
section~\ref{accel}. Since time is measured in units of $\tau$ in such
a scheme, the curve thus found is the limiting master curve that would
have been found by the conventional means of \cite{glKawa} in the
limit $T\to0$. We indeed checked that the amplitude of the
law~(\ref{Lt}), $L(t)=A\, (t/\tau)^{1/3}$, found in~\cite{glKawa} was
in agreement, though less precise, than that found in the present
work, $A \simeq 2.29$. Figure~\ref{f1} depicts the master curve thus
obtained in the $T\to0$ limit. The system size is $N=2^{12}=4096$, and
the number of samples are reported in Table~\ref{stat}. (We also
checked for the absence of any finite-size effects by simulating few
$N=2^{16}$ samples.)

\begin{table}
\begin{tabular}{|c|c|c|c|c|c|c|}
\hline
\# evap & 0 & $10^3$ & $10^4$ & $10^5$ & $10^6$ & $10^7$ \\
\hline
$s$ value & $0^+$ &\ 1.1710(5)\ \ &\ 15.998(12)\ \ &\ 312.43(23)\ \ &
\ 8000(9)\ \ &\ 235580(740)\ \ \\
\hline
\ \# samp\ \ &\ 895\ \ & 1189 & 717 & 1146 & 1179 & 51 \\
\hline
\end{tabular}
\caption{For each choice of the number of evaporation processes done
before the measurement of the autocorrelation and response, the
corresponding value of the waiting time $s$ and the number of
different thermal histories are reported.}
\label{stat}
\end{table}

The inset of figure~\ref{f1} shows the growth of $L(t)$ in a log-log
scale. The convergence to the slope 1/3 is slow, mainly because of the
presence of an offset at initial times. This offset corresponds to the
first blocked state reached by the system. Remind that the system is
prepared at time $t=0$ in a random configuration.  It is then evolved
under $T=0$ Kawasaki dynamics, until it reaches the first blocked
state at time $t=0^+$ in units of $\tau$.  This first blocked state
has a mean domain length $L(0^+) = 4.15886(15)$.

\subsection{Autocorrelation}

A well-known fact of the kinetics of coarsening with non-conserved
dynamics is that, in the low-temperature scaling regime, and for large
temporal separations ($1\ll s\ll t\ll t_{\rm eq}$), the two-time
autocorrelation $C(t,s)$ decays as~\cite{bray}
\be
C(t,s)\sim\left(\frac{L(t)}{L(s)}\right)^{-\lambda},
\label{1}
\ee
and, for the particular case where $s=0$, as
\be
C(t,0)\sim L(t)^{-\lambda},
\label{2}
\ee
defining the autocorrelation exponent $\lambda$~\cite{husefisher}.

The case of conserved dynamics is more complex. For quenches to
temperatures below $T_c$, Yeung et al.~\cite{yeung} find bounds on the
autocorrelation exponent which depend on the value of the smallest
time $s$. For $s=0$, $\lambda\ge d/2$, where $d$ is the dimensionality
of space, while for values of $s$ in the scaling regime, $\lambda\ge
d/2+2$ for $d\ge 2$, and $\lambda\ge3/2$ for $d=1$.

Given the prediction of ref.~\cite{maj} that,
in the low-temperature phase, $C(t,0)\sim
L(t)^{-\lambda}$, with $\lambda=d$, Yeung et al. conclude that, for
$d=1$ (and more generally for low dimensions), the
behaviour~(\ref{2}) holds for $s$ small, while for $s$ in the scaling
regime the behaviour~(\ref{1}) should be replaced by
\be
C(t,s)\sim\left(\frac{L(t)}{L(s)}\right)^{-\lambda'}
\label{3}
\ee
with a different exponent $\lambda'>\lambda$.

The prediction above therefore implies that for
the Ising-Kawasaki chain at vanishingly small temperature
the curve of the
autocorrelation for two values of $s$, one being taken in the
short-time regime, the other one in the scaling regime, should cross
at some later time $t^*$. This simple observation was not noticed by
the authors of~\cite{yeung}. The question therefore arises of what is
the behaviour of the autocorrelation $C(t,s)$ for later times $t\gg
t^*$.

Hereafter we suggest the following scenario, based on the reasonable
hypothesis that the two curves mentioned above do not cross. Define
more precisely $t^*(s)$ by
\be
L(t^*)^{-\lambda}\sim
\left(\frac{L(t^*)}{L(s)}\right)^{-\lambda'},
\ee
i.e.\ $t^*(s)\sim s^{\lambda'/(\lambda'-\lambda)}$. Then, there are
two scaling regimes:
\begin{itemize}
\item the intermediate scaling regime of Yeung et al., for $t\ll t^*$,
where~(\ref{3}) holds,
\item the ultimate scaling regime, for $t\gg t^*$, where
\be
C(t,s)\sim L(t)^{-\lambda}.
\label{4}
\ee
\end{itemize}
Therefore the following scaling law should hold
\be
C(t,s)=L(t)^{-\lambda}g\left(\frac{L(t)^{\lambda'-\lambda}}
{L(s)^{\lambda'}}\right),
\label{g}
\ee
with $g(x\to0)\sim x^{-1}$ in the intermediate regime, while in the
ultimate regime $g(x\to\infty)$ should converge to a constant.

\begin{figure}[htb]
\begin{center}
\includegraphics[width=.75\linewidth]{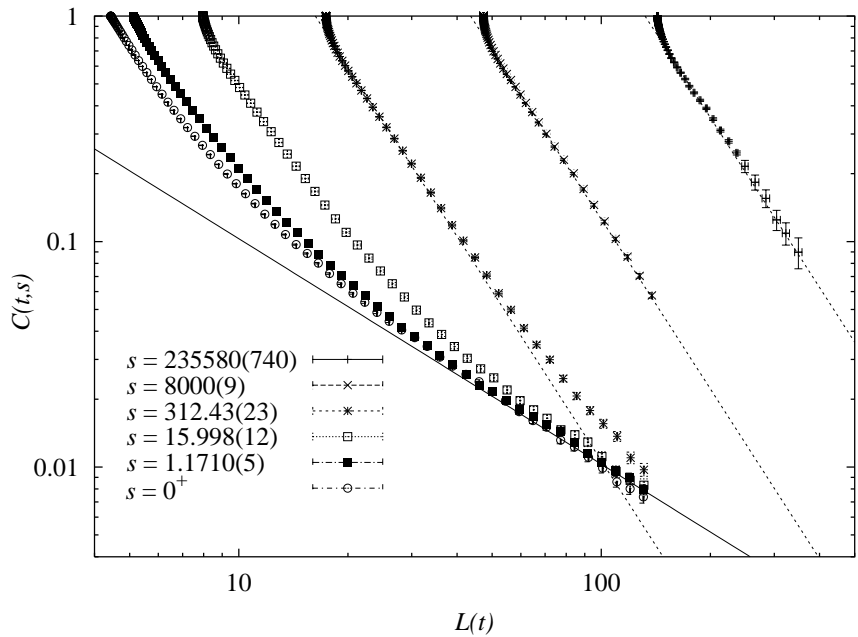}
\caption{Two-time autocorrelation function of the Ising-Kawasaki chain
at vanishingly small temperature. Full line has slope $-\lambda=-1$,
while dotted ones have slope $-\lambda'=-2.5$.}
\label{corr_L}
\end{center}
\end{figure}

\begin{figure}[htb]
\begin{center}
\includegraphics[width=.75\linewidth]{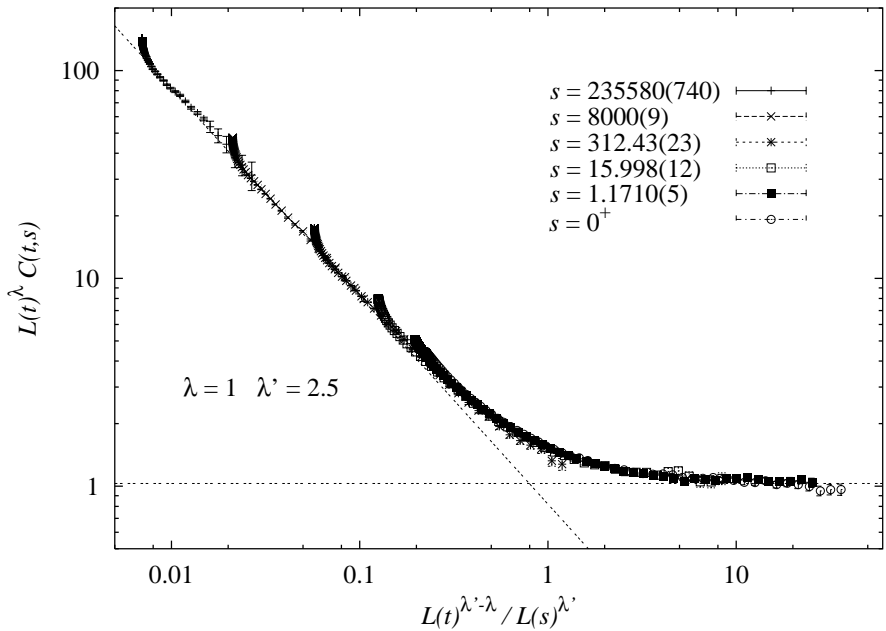}
\caption{Scaling function for the two-time autocorrelation of the
Ising-Kawasaki chain at vanishingly small temperature (see text).}
\label{f2}
\end{center}
\end{figure}

Outcomes from our simulations are compatible with these
predictions. Figure~\ref{corr_L} shows the autocorrelation function as
a function of $L(t)$, for the values of $s$ given in Table~\ref{stat}.
The presence of two different scaling regimes is evident, although the
intermediate regime is not very clean on small $s$ data, and the
ultimate regime is not reached for large $s$ data. Figure~\ref{f2} is
a plot of the scaling function $g(x)$ defined above. We find $\lambda'
\simeq 2.5$, with $\lambda=1$.

\begin{figure}[htb]
\begin{center}
\includegraphics[width=.75\linewidth]{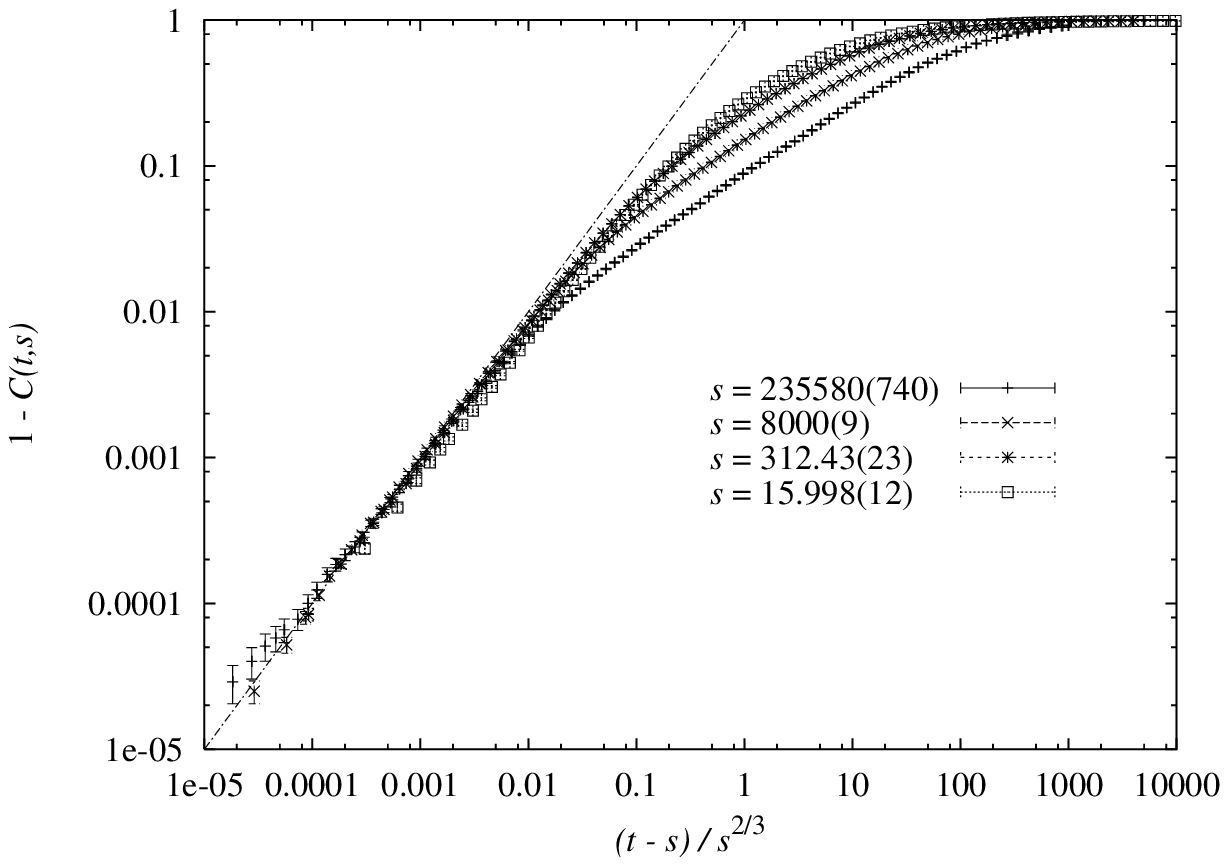}
\caption{Scaling of the autocorrelation function at very short times.}
\label{short_times}
\end{center}
\end{figure}

Let us note that before attaining the intermediate scaling regime
there is yet another temporal regime, clearly visible on
figures~\ref{corr_L} and \ref{f2}, which takes place at very short
times.  One can show that in the scaling variable of figure~\ref{f2}
this very short time regime becomes smaller when $s$ increases, and
therefore irrelevant in the $s \to \infty$ limit. Indeed one finds
numerically --see figure~\ref{short_times}-- that in the very short
time regime the autocorrelation function can be written as follows
\be
1 - C(t,s) \approx \frac{t-s}{s^{2/3}} \quad .
\label{eq_short_times}
\ee
Eq.~(\ref{eq_short_times}) implies that the timescale of decorrelation
processes happening in the very short time regime is $s^{2/3}$. This
timescale is much smaller than the timescale of the intermediate
regime, which is $s$.

In preparation of section~\ref{2dcrit}, let us mention that, at
criticality, dynamical scaling predicts that
\be
C(t,0) \sim L(t)^{-\lambda_c},
\label{crit0}
\ee
defining the critical autocorrelation exponent
$\lambda_c$~\cite{huse1,janssen}. We have also, for $s$ in the 
scaling regime,
\be
C(t,s) \sim L(s)^{-2\beta/\nu}\left(\frac{L(t)}{L(s)}\right)^{-\lambda_c}
\label{crits}
\ee
where $\beta$ and $\nu$ are the usual static critical exponents
($\beta=1/4$, $\nu=1$, in 2D). This form holds for both non-conserved
and conserved dynamics, with the possibility that, for the latter, the
exponent appearing in (\ref{crits}) be not the same as in
(\ref{crit0}), as discussed in section~\ref{2dcrit}.  For conserved
dynamics at criticality, refs.~\cite{maj,alexander,satya} predict
$\lambda_c=d$, while statements made in \cite{yeung} on the long-time
behaviour of the autocorrelation are less precise.

A last comment is in order.  At criticality, there are no well-defined
growing domains.  The interpretation of the growing length in
eqs.~(\ref{crit0}) and (\ref{crits}) is the typical size over which
the system looks critical.  The present situation of a system evolving
after a quench from high temperature to $T=T_c$ is nevertheless
usually referred to as ``critical coarsening''.  By convenience we
shall still call the length $L(t)$ the mean domain size.

Note that for the Ising chain the magnetization exponent $\beta=0$,
hence there is no distinction to be made between the two behaviours
~(\ref{1}) and (\ref{crits}).

\subsection{Response and fluctuation-dissipation plot}

Let us recall that for the zero-temperature non-conserved (Glauber)
case, the two-time correlation~\cite{bray} and response functions are
known analytically~\cite{gl1D,lip}. As a consequence, the
fluctuation-dissipation ratio $X(t,s)$, defined by~\cite{cuku2}
\[
R(t,s)=\frac{X(t,s)}{T}\frac{\partial C(t,s)}{\partial s},
\]
can be obtained in closed form. In the scaling regime, it reads
\[
X(t,s)\approx
\frac{1}{2}\left(1+\frac{s}{t}\right),
\]
yielding, for large temporal separations, the limiting ratio
\[
X_{\infty }=\lim_{s\to\infty}\lim_{t\to\infty}X(t,s)=\frac{1}{2}.
\]
In the scaling regime, the integrated response functions $\rho(t,s)$
and $\chi(t,s)$ only depend on $t/s$, or equivalently on $C$,
\be
\rho(C)=\frac{1}{\sqrt{2}}-\chi(C)=
\frac{\sqrt{2}}{\pi }\arctan \left( \frac{1}{\sqrt{2}}\tan
\frac{\pi C}{2}\right) .
\label{fdplot}
\ee
Alternatively the limiting ratio can be extracted from $\rho(C)$, as
\be
X_{\infty }=\lim_{C\rightarrow 0} \frac{\rho(C)}{C}.
\label{xinfty}
\ee
Note that, in contrast to the generic two-dimensional case considered
in section~\ref{2dcrit}, for the Ising chain, the integrated response
function does not bear any dependence in $s$ because the exponent
$\beta =0$.

\begin{figure}[htb]
\begin{center}
\includegraphics[width=.75\linewidth]{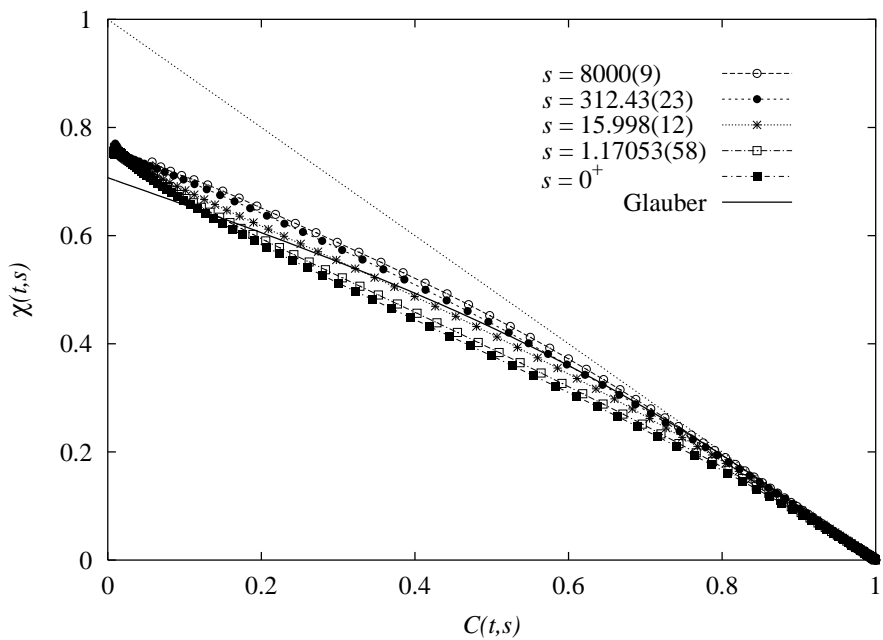}
\caption{Integrated response against autocorrelation for the
Ising-Kawasaki chain at vanishingly small temperature.}
\label{f3}
\end{center}
\end{figure}

We now turn to the Ising-Kawasaki chain. Figure~\ref{f3} depicts a
plot of the integrated response $\chi(t,s)$ against the correlation
$C(t,s)$, in the limit of vanishingly small temperature, for the
values of the waiting time $s$ given in Table~\ref{stat}.  Most of the
plot corresponds to the intermediate regime ($t\ll t^*(s)$), since the
ultimate regime is reached only for the two smallest $s$ values and
very small correlations $C(t,s) \alt 0.03$ (see figure~\ref{corr_L}).
Restricting to the first regime, we observe a slow convergence of the
data to a limiting curve when $s$ increases. However this limiting
curve lies above the theoretical Glauber curve~(\ref{fdplot}), in
contradiction with the prediction made in ref.~\cite{salerno1},
stating that the fluctuation-dissipation plot for the Ising-Kawasaki
chain is identical to that of the Ising-Glauber chain, for vanishingly
small temperature.

We are thus led to critically review ref.~\cite{salerno1}. A similar
plot of the integrated response against correlation is presented in
this reference, for three sets of values of $T$ and $s$:
$(T=0.48,s=3\cdot 104)$, $(T=0.70, s=2\cdot 103$), and
$(T=0.70,s=104)$.  For these values of $(T,s)$ the data follow rather
closely, at least in a range of values of the correlation, the
theoretical Glauber curve~(\ref{fdplot}). The authors of
ref.~\cite{salerno1} therefore induce that the data obtained for
vanishingly small temperatures and increasing values of $s$ should
eventually converge to eq.~(\ref{fdplot}).

Noting that the values of $(T,s)$ mentioned above correspond, in units
of $\e^{4\beta}$, to $s=7.2$, $s=6.6$, and $s=33$ respectively, we see
on figure~\ref{f3} that, indeed, for this range of values of $s$, the
data points fall not too far away from the Glauber curve. However,
since this holds neither for smaller nor for larger values of $s$ (in
units of $\e^{4\beta}$), we conclude that the apparent identity
between Kawasaki and Glauber curves observed in~\cite{salerno1} is
just a coincidence due to the range of values considered in this
reference.

The existence, in the intermediate regime, of a non trivial scaling
limit for the response implies, as a corollary, a non trivial value of
$X_\infty$, when the ratio $x=t/s\to\infty$.  A precise numerical
value of this quantity is however difficult to obtain from ZFC
data. At present we can not exclude the possibility that $X_\infty$ is
the same for both Glauber and Kawasaki dynamics.

Consider now the ultimate regime ($t\gg t^*(s)$).  This regime
corresponds to $C<\widetilde{C}^*(s)$ where $\widetilde{C}^*(s)\equiv
C(t^*(s),s)\sim s^{-\lambda \lambda'/(\lambda'-\lambda)}$, with
$\lambda \lambda'/(\lambda'-\lambda)\simeq 1.66$.  This is reminiscent
of the situation encountered at criticality for the two-dimensional
Ising-Glauber system~\cite{gl2D}. In this case the
fluctuation-dissipation theorem holds except in a region $C<C^*(s)\sim
s^{-a_c}$, where $a_c=2\beta/\nu z_c\simeq 0.115$, which vanishes for
increasing values of $s$. However this mechanism does not prevent the
occurrence of a non trivial limiting ratio
$X_\infty$~\cite{gl2D}. (See section~\ref{2dcrit}.)  By analogy, we
expect a non trivial value of $X_\infty$ in the ultimate regime, a
priori different from that obtained in the intermediate regime.  Note
however that since the exponent $\lambda \lambda'/(\lambda'-\lambda)$
is larger than 1, $\widetilde{C}^*(s)$ is decreasing very fast, and,
as a consequence, the regime where $X_\infty$ could be measured is
hardly reachable in a numerical simulation.

\section{Results for the critical dynamics of the 2D Ising-Kawasaki
model}
\label{2dcrit}

The aim of this section is to investigate the critical coarsening of a
two-dimensional system of spins evolving under Kawasaki dynamics from
a random initial condition. We will, as in the previous section, and
inspired by the results found there, examine in turn the behaviour of
the two-time correlation function, then of the two-time response
function.

\subsection{Mean domain length}

The numerical study of the long-time behaviour of $C(t,s=O(1))$ for
the critical dynamics of the 2D Ising-Kawasaki model is well
established~\cite{alexander}. Scaling is observed, i.e.\
eq.~(\ref{crit0}) holds, with a decay exponent $\lambda_c = 2$,
confirming the prediction of~\cite{maj}.  The mean domain size itself
is observed to grow as $L(t)\sim t^{1/z_c}$, with
$z_c=4-\eta=15/4$~\cite{alexander}. A confirmation of these results is
provided by figure~\ref{f4}, where the mean size of domains is
extracted by the excess energy with respect to the equilibrium energy
at $T_c$, $E_{\rm eq}=1/\sqrt{2}$, i.e.\ $L(t)=(E(t)-E_{\rm
eq})^{-1}$.  This method has been already used for low temperature
coarsening, either for conserved~\cite{amar}, or non conserved
dynamics~\cite{husefisher}.  It is however the first time that it is
used at criticality, where in general the growing length scale $L(t)$
is obtained from the position of the first zero of the equal-time
correlation function.  Our reason to do so lies in the fact that for
critical coarsening dynamical scaling holds, as it does for low
temperature coarsening, and that therefore there exists only one
single growing length scale in the system.

\begin{figure}[!htb]
\begin{center}
\includegraphics[width=.75\linewidth]{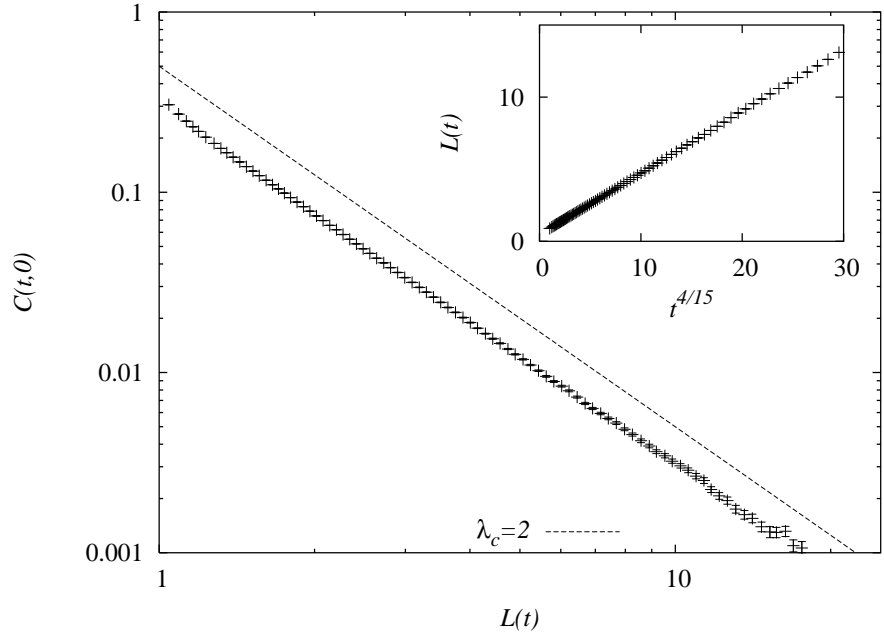}
\caption{2D Ising-Kawasaki model at criticality: $C(t,0)$ and
$L(t)$. }
\label{f4}
\end{center}
\end{figure}
\begin{figure}[!htb]
\begin{center}
\includegraphics[width=.495\linewidth]{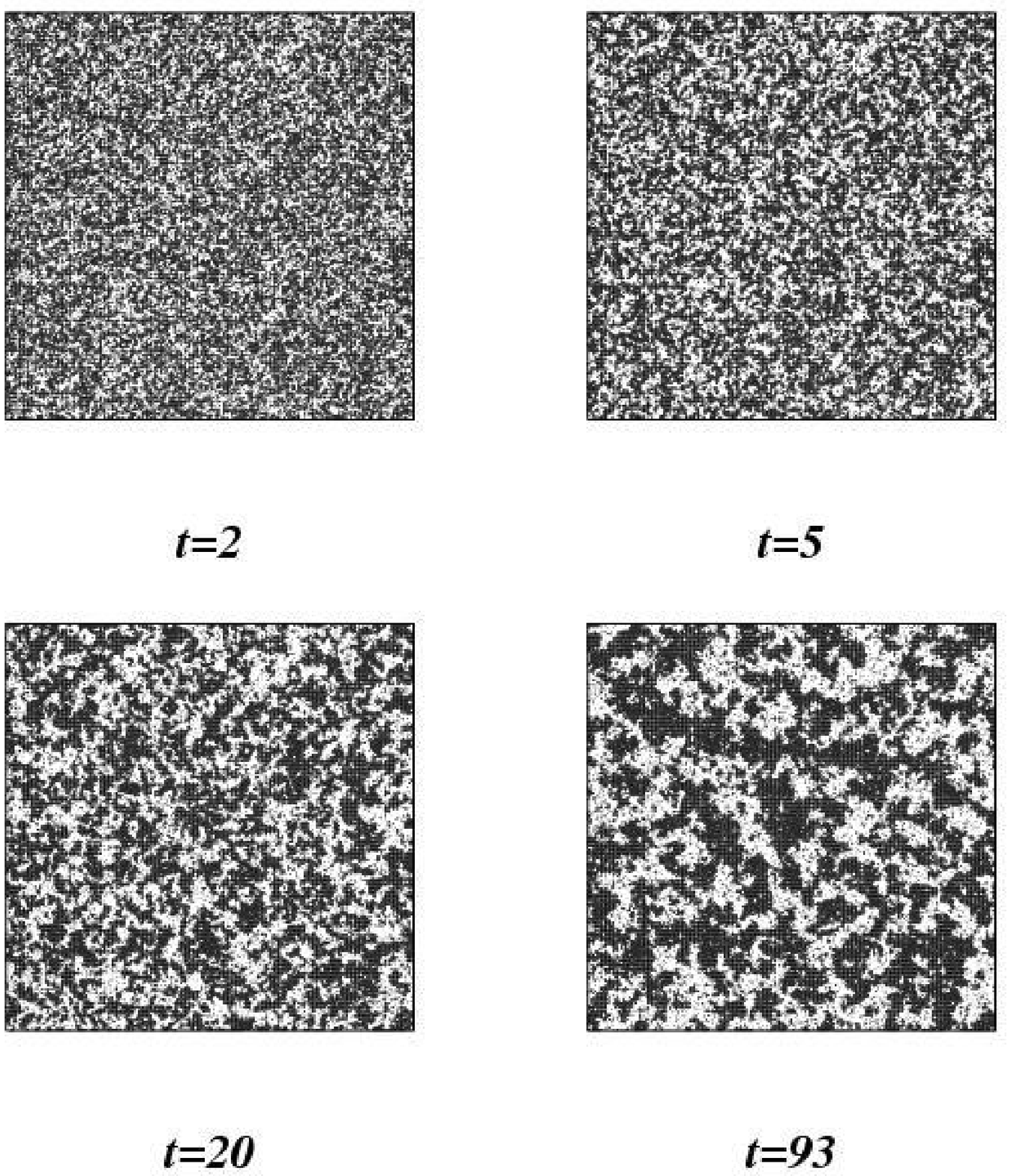}
\hfill
\includegraphics[width=.495\linewidth]{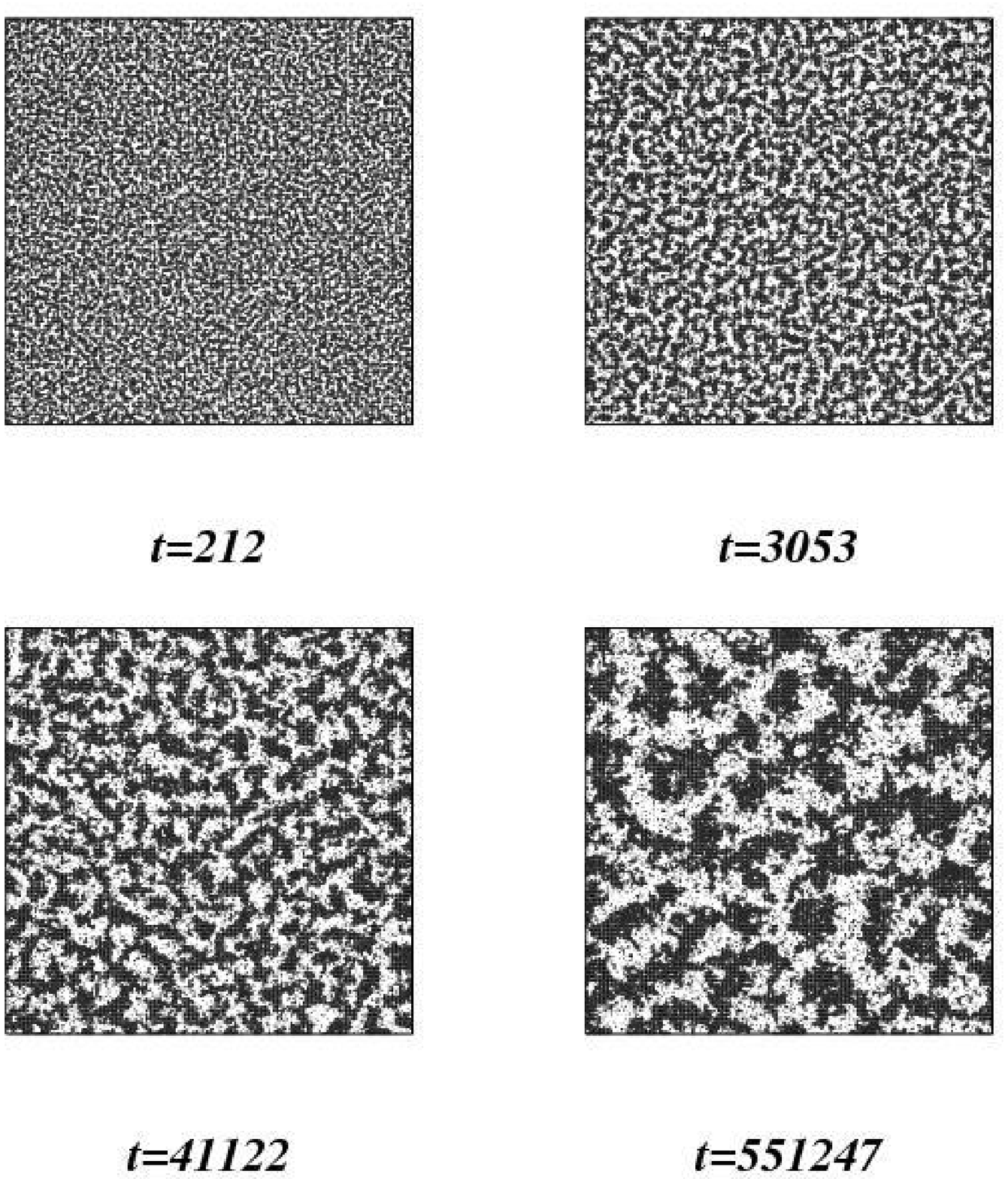}
\caption{Snapshots of configurations after a quench from $T=\infty$ to
$T=T_c$ in a $500^2$ Ising spin system. The four snapshots on the left
correspond to Glauber dynamics, while the four on the right to
Kawasaki dynamics. Times have been chosen in order to have similar
values for $L(t)$.}
\label{snap}
\end{center}
\end{figure}

In order to compare the qualitative behaviour of both conserved and
non-conserved dynamics at $T_c$, we take a series of snapshots at
instants of times where similar domain sizes were reached in both
dynamics (see figure~\ref{snap}).  Since for non-conserved dynamics
$z_c\approx 2.17$, in 2D, conserved dynamics is much slower.  It is
interesting to note the overall similarity of the snapshots at
corresponding instants of time.

\subsection{Autocorrelation}

The study of the behaviour of $C(t,s)$ when the waiting time $s$ is
deep in the scaling regime is largely unexplored.  We first measured
the autocorrelation as a function of $L(t)$ for different values of
$s$ (see fig.~\ref{f5}).  As in the one-dimensional case, we observe a
seemingly different decay exponent as soon as $s$ is large enough,
i.e.\ in this regime eq.~(\ref{crits}) should be read with a different
exponent $\lambda'_c$.

\begin{figure}[htb]
\begin{center}
\includegraphics[width=.75\linewidth]{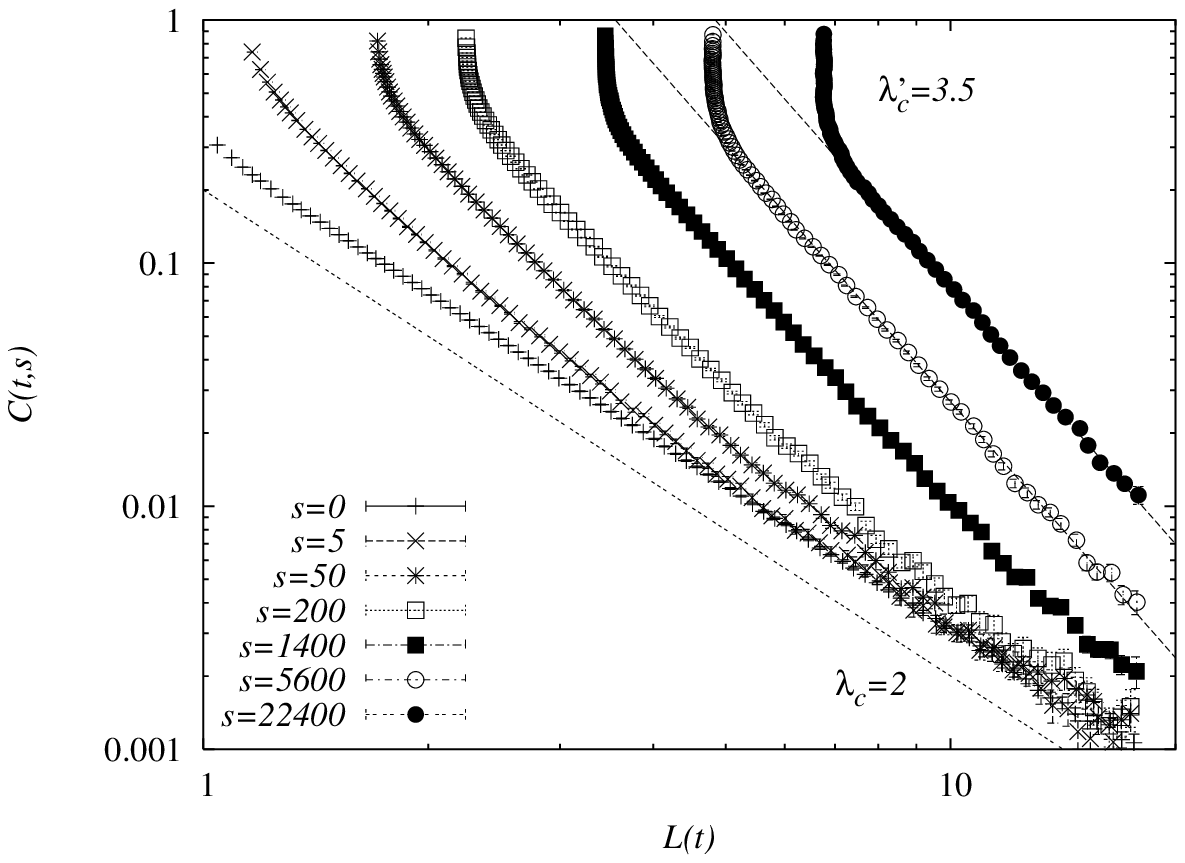}
\caption{2D Ising-Kawasaki model at criticality: $C(t,s)$ against
$L(t)$ for different values of $s$.}
\label{f5}
\end{center}
\end{figure}
\begin{figure}[htb]
\begin{center}
\includegraphics[width=.75\linewidth]{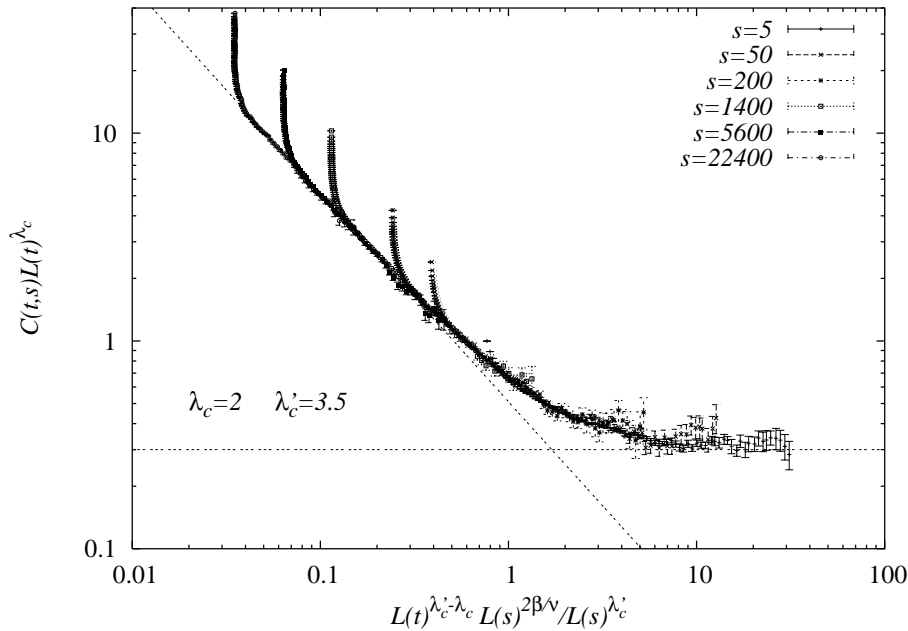}
\caption{2D Ising-Kawasaki model at criticality: scaling function for
the two-time autocorrelation.}
\label{f6}
\end{center}
\end{figure}

In order to assess this point, we define, by analogy with the
one-dimensional case (see eq.~(\ref{g})), the scaling function $g_c$
as
\be
C(t,s)=L(t)^{-\lambda_c} g_c\left(\frac{L(t)^{\lambda'_c-\lambda_c}}
{L(s)^{\lambda'_c-2\beta/\nu}}\right),
\ee
with $g_c(x\to0)\sim x^{-1}$ in the intermediate regime, while in the
ultimate regime $g_c(x\to\infty)$ should converge to a constant.
Figure~\ref{f6} depicts the scaling function obtained using
$\lambda_c=2$ and $\lambda'_c=3.5$.  In view of this figure it is
reasonable to conclude again in favour of the existence of these two
scaling regimes, defined by the relative magnitude of $t$ with respect
to the crossover timescale $t^*(s) \sim
s^{(\lambda'_c-2\beta/\nu)/(\lambda'_c-\lambda_c)}$.  Otherwise
stated, $t\ll t^*(s)$ in the intermediate regime, while $t\gg t^*(s)$
in the ultimate regime.  We have, with the values of the exponents
given above, and $2\beta/\nu=1/2$, $t^*(s) \sim s^2$.

\subsection{Response and fluctuation-dissipation plot}

We now turn to the response.  Following~\cite{gl2D}, we choose to
compute the thermoremanent magnetization~(\ref{trm}) (TRM).  Due to
the fast increase of the crossover timescale, $t^*(s) \sim s^2$, and
to the extreme difficulty to have precise data for the response at
very long times, the results presented below only concern the
intermediate regime.  These results are contained in figures~\ref{f7},
~\ref{f7_b}, and \ref{f8}.

In the intermediate regime we assume for the TRM a scaling form
similar to that of the correlation function, equation~(\ref{crits}),
that is
\bea
\rho(t,s) \sim L(s)^{-2\beta/\nu}
\left(\frac{L(t)}{L(s)}\right)^{-\lambda'_c}.
\label{rho_scal}
\eea
%
\begin{figure}[htb]
\begin{center}
\includegraphics[width=.495\linewidth]{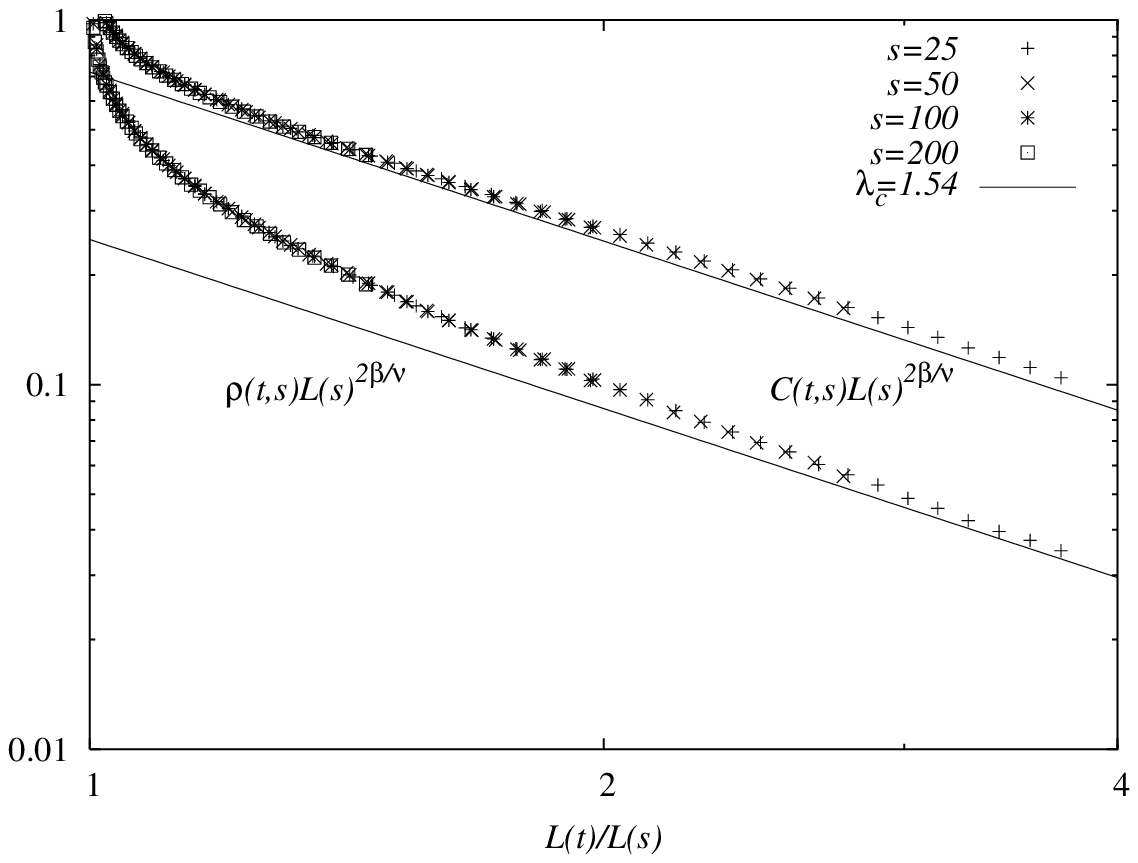}
\includegraphics[width=.495\linewidth]{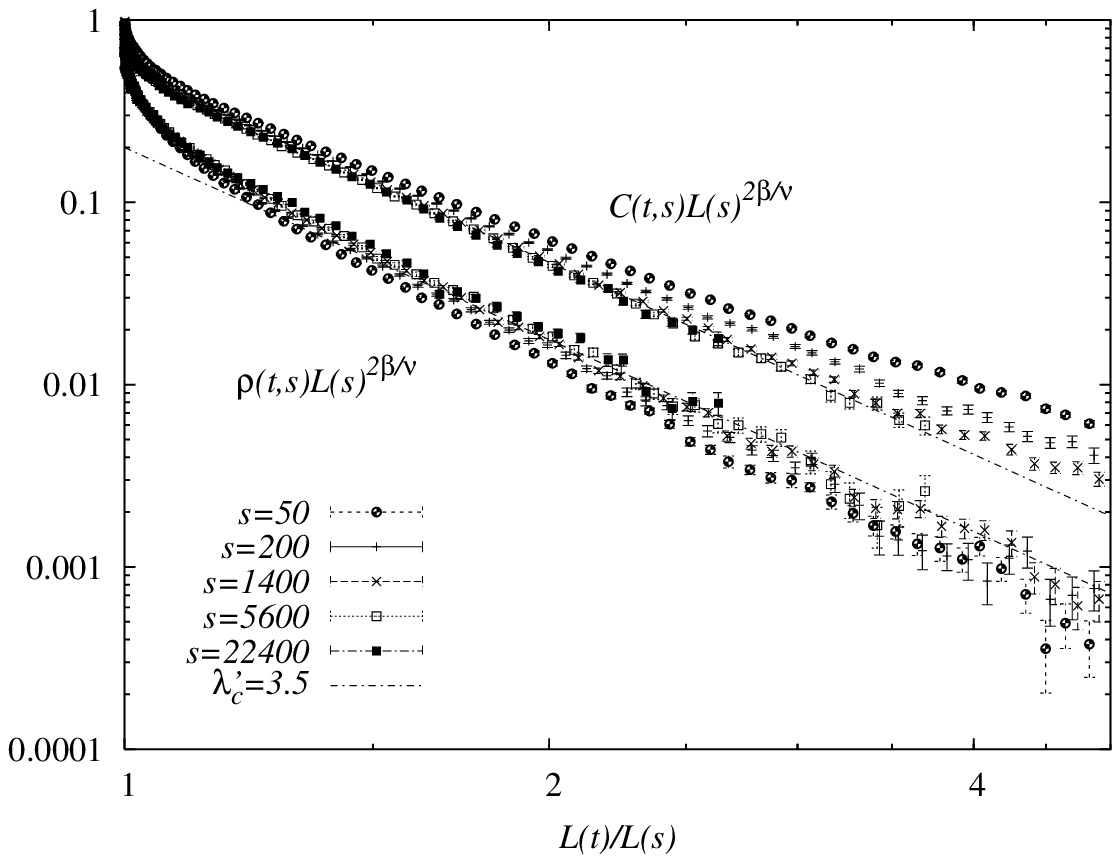}
\caption{2D Ising model at criticality: scaling functions of
correlation and response.  (left) Glauber dynamics.  The full lines
have slope $\lambda_c=-1.54$.  (right) Kawasaki dynamics.  The dotted
lines have slope $\lambda'_c=-3.5$.  }
\label{f7}
\end{center}
\end{figure}
This scaling form is well verified by our numerical data, as
illustrated by figure~\ref{f7}, which depicts plots of the rescaled
correlation and response functions, both for Glauber and Kawasaki
dynamics.  For the latter, it is interesting to note that, when $s$
increases, the master curve is attained from above for the
correlation, and from below for the response, indicating that
asymptotically the two scaling functions should have the same
algebraic decay with exponent $\lambda'_c$.  Plots of the scaling
functions of autocorrelation and response for the two-dimensional
Ising-Glauber model at criticality first appeared in~\cite{gl2D}.  The
value of the exponent measured in the present work is slightly smaller
than that found in~\cite{gl2D}.

\begin{figure}[htb]
\begin{center}
\includegraphics[width=.75\linewidth]{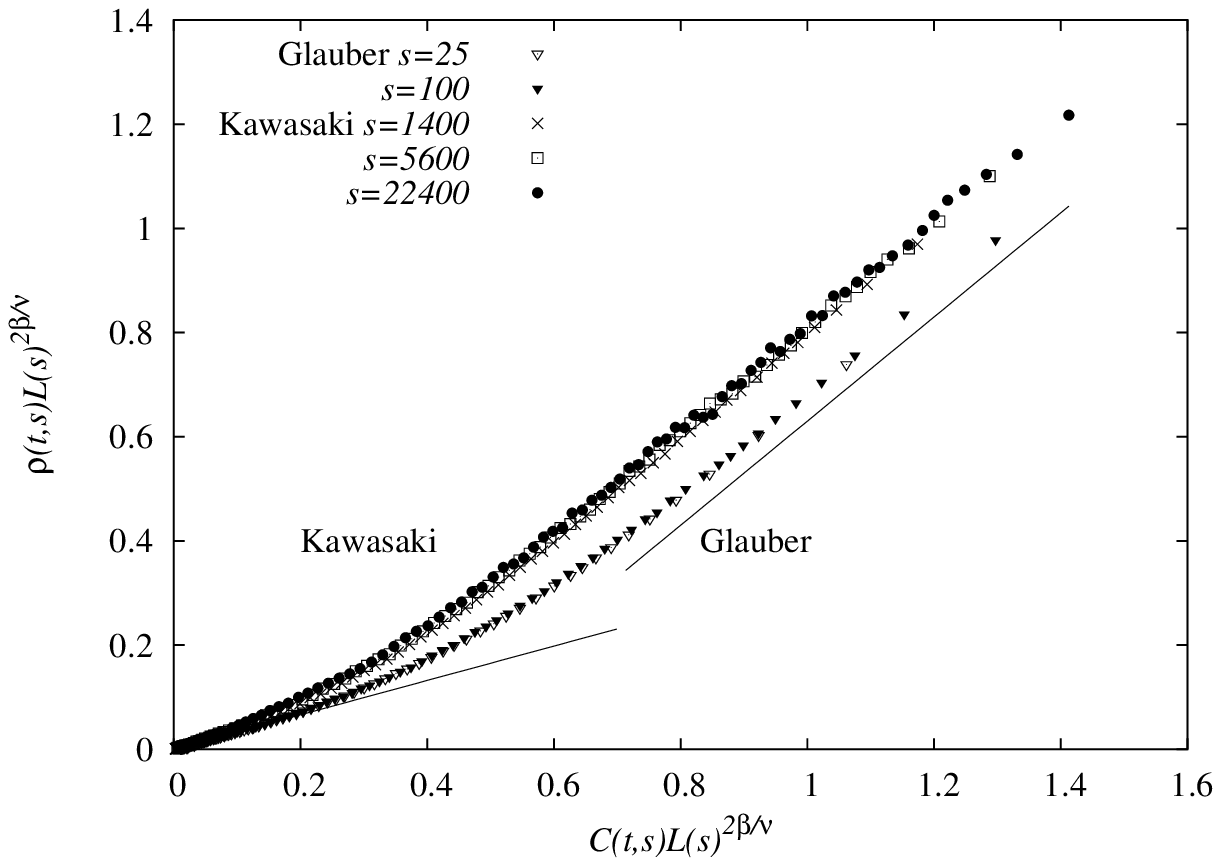}
\caption{2D Ising model at criticality: parametric plots of rescaled
response versus rescaled correlation for Glauber and Kawasaki
dynamics, using the data of figure~\ref{f7}.  The full line at the
origin has slope $0.33$.  The other one has slope 1, and is meant as a
guide to the eye.}
\label{f7_b}
\end{center}
\end{figure}

Another representation of the same data is given in figure~\ref{f7_b},
which depicts the parametric plot of the rescaled response against the
rescaled correlation, both for Glauber and Kawasaki dynamics.  This
figure shows that the two dynamics lead to the same phenomenology.
Indeed, define a crossover scale for the autocorrelation by
$C^*(s)=C(2s,s)$~\cite{gl2D}, corresponding, for $s$ large enough, to
a value of the abscissa on figure~\ref{f7_b} approximately equal to
$0.47$ for Glauber dynamics, and to $0.27$ for Kawasaki dynamics.
Then, on the right part of the plots with respect to these values, the
slope of the two curves is equal to one, in agreement with the
fluctuation-dissipation theorem.  The left part of the plots
corresponds to the scaling regime, with a crossover towards a non
trivial slope at the origin, equal to the limiting violation ratio
$X_\infty$.

\begin{figure}[htb]
\begin{center}
\includegraphics[width=.495\linewidth]{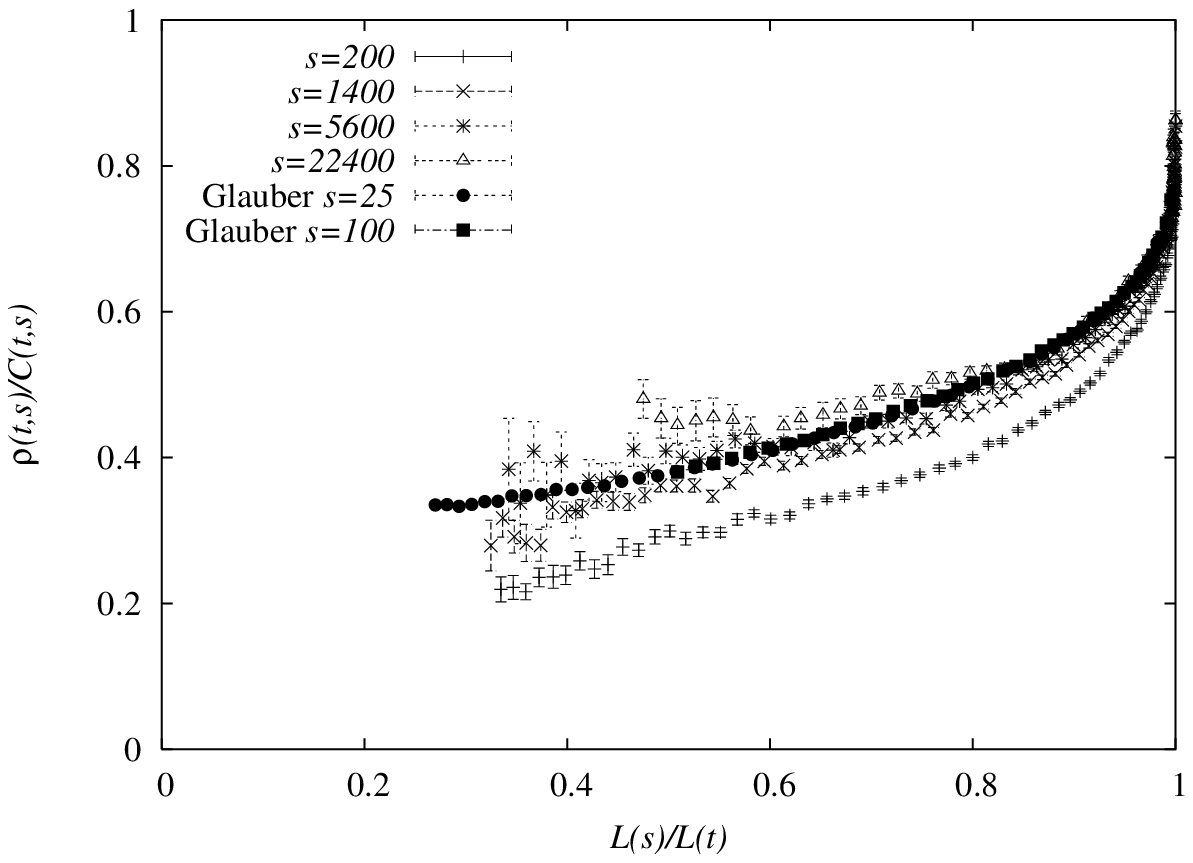}
\includegraphics[width=.495\linewidth]{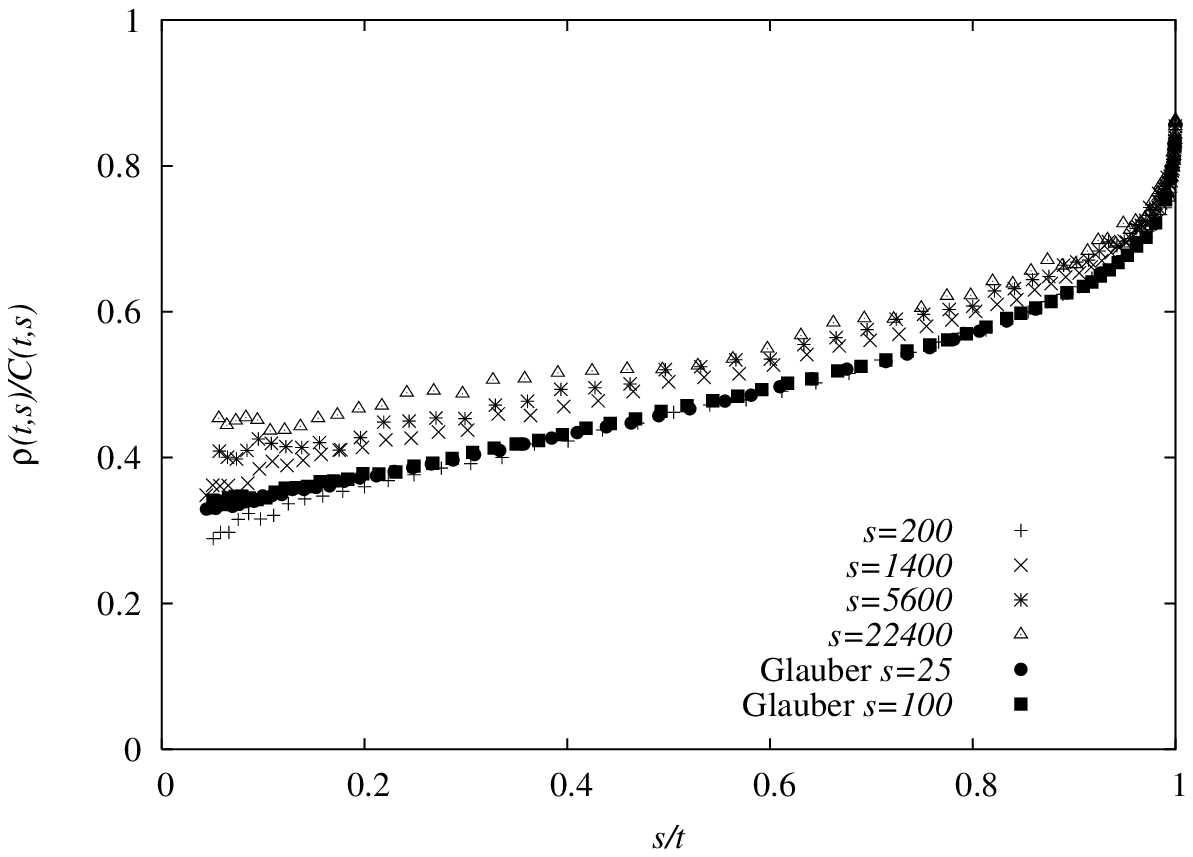}
\caption{2D Ising model at criticality: (left) $\rho/C$ versus
$L(s)/L(t)$; (right) $\rho/C$ versus $s/t$.}
\label{f8}
\end{center}
\end{figure}
In order to extract the numerical estimate of these slopes from these
data, we plot them in two different fashions, as shown on
figure~\ref{f8}.  Since $\rho(t,s)\approx X_\infty\,C(t,s)$ (see
(\ref{xinfty})), we plot the ratio ${\rho}/{C}$, first versus
$L(s)/L(t)$ (figure~a), then versus the ratio of times $s/t$
(figure~b). Error bars are not shown in the latter, in order to
improve readability. For Glauber dynamics $X_\infty$ seems to be
attained linearly in $s/t$ , which leads, by taking the intercept with
the $y$-axis, to the prediction $X^{\rm Glauber}_\infty \approx 0.33$,
in agreement with the estimates given in recent
works~\cite{mayer1,chatelain}.  Note that for the spherical model the
correction to $X_\infty$ is exactly in $s/t$~\cite{gl2D}.  Finally,
though it is difficult to be conclusive on the sole basis of
figure~\ref{f8}, the latter gives more evidence in favour of different
values for the limiting ratios corresponding to the two dynamics, with
$X^{\rm Kawasaki}_{\infty} > X^{\rm Glauber}_{\infty}$.

\section{Conclusion}

Let us summarize the most salient points of this study.

We extend the method of~\cite{chatelain,fede} to the case of Kawasaki
dynamics.  We introduce a new method for the investigation of the
low-temperature scaling regime of the Ising-Kawasaki chain.  We define
rules for an accelerated dynamics, which are both faithful and
efficient.  This method can be extended to higher dimensions.  We find
new results concerning the behaviour at large time of the
autocorrelation function for the critical Kawasaki dynamics in both
dimensionalities, and demonstrate the existence of an ultimate regime,
which was overlooked in previous studies.  We believe this regime to
be also present in the behaviour of the response, though hardly
accessible with present computer capabilities.  As a corollary, we
expect the existence of a different value of the limiting ratio
$X_\infty$ in the ultimate regime, which would require time scales
which are presently unreachable.

In the course of this study we were led to question the validity of
the results of ref.~\cite{salerno1} concerning the
fluctuation-dissipation plot for the Ising-Kawasaki chain.  The
evidence, claimed in ~\cite{salerno1}, for the identity of the
fluctuation-dissipation plots for Glauber and Kawasaki dynamics is
seemingly coincidental, and lies in the range of values used in this
reference.

For the two-dimensional system at criticality, the
fluctuation-dissipation plots obtained from the two dynamics are
phenomenologically similar, but not quantitatively identical.  In
particular, the limiting violation ratios $X_\infty$ are found to be
different.  It is harder though to be conclusive in the
two-dimensional case than in the one-dimensional one.  In the former,
the scaling region defined by $C(t,s)<C^*(s)\sim s^{-0.115}$ is small,
while in the latter, since $\beta=0$, it identifies to the whole range
$0\le C(t,s)\le1$.

In the intermediate scaling regime numerical data are compatible with
the simple form $\lambda'_c=d + 3/2$ for the autocorrelation exponent.
It would be interesting to assess the validity of this hypothesis.  A
theoretical explanation of the existence of the ultimate regime, both
for the autocorrelation and for the response, is needed.  Finally, a
natural extension of the present work is the study of the behaviour of
the autocorrelation and response in the low-temperature phase
($T<T_c$) of the two-dimensional kinetic Ising model with Kawasaki
dynamics.

\begin{acknowledgments}
C.G. wants to thank Giorgio Parisi for his warm hospitality at the
Statistical Mechanics and Complexity Center, where this work was
initiated.  F.K. and F.R.-T. acknowledge the financial support
provided through the European Community's Human Potential Programme
under contracts HPRN-CT-2002-00319, Stipco and HPRN-CT-2002-00307,
Dyglagemem.
\end{acknowledgments}


\begin{thebibliography}{}

\bibitem{bray} A.J. Bray, Adv. Phys. {\bf 43}, 357 (1994).

\bibitem{janssen} H.K. Janssen, B. Schaub, and B. Schmittmann,
Z. Phys. B {\bf 73}, 539 (1989).

\bibitem{huse1} D.A. Huse, Phys. Rev. B {\bf 40}, 304 (1989).

\bibitem{cuku1} L.F. Cugliandolo, J. Kurchan, and G. Parisi,
J. Physique I {\bf 4}, 1641 (1994).

\bibitem{cuku2} L.F. Cugliandolo and J. Kurchan J. Phys. A
\textbf{27}, 5749 (1994).

\bibitem{barrat} A. Barrat Phys. Rev. E \textbf{57}, 3629 (1998).

\bibitem{berthier} L. Berthier, J. L. Barrat, and J. Kurchan
Eur. Phys. J. B \textbf{11}, 635 (1999).

\bibitem{gl1D} C. Godr\`eche and J.M.~Luck, J. Phys. A \textbf{33},
1151 (2000).

\bibitem{lip} E. Lippiello and M. Zannetti, Phys. Rev. E {\bf 61},
3369 (2000).

\bibitem{gl2D} C. Godr\`eche and J.M.~Luck, J. Phys. A \textbf{33},
9141 (2000).

\bibitem{malte} M. Henkel, M. Pleimling, C. Godr\`eche, and J.M.~Luck,
Phys. Rev. Lett. {\bf 87}, 265701 (2001).

\bibitem{cala} P. Calabrese and A. Gambassi, Phys. Rev. E {\bf 65},
066120 (2002).

\bibitem{calb} P. Calabrese and A. Gambassi, Phys. Rev. B {\bf 66},
212407 (2002).

\bibitem{pico} A. Picone and M. Henkel, J. Phys. A {\bf 35}, 5575
(2002).

\bibitem{mayer1} P. Mayer, L. Berthier, J.P. Garrahan, and P. Sollich,
Phys. Rev. E {\bf 68}, 016116 (2003).

\bibitem{mayer2} P. Mayer and P. Sollich, J. Phys. A {\bf 37}, 9
(2004).

\bibitem{cris} A. Crisanti and F. Ritort, J. Phys. A {\bf 36}, R181
(2003).

\bibitem{sastre} F. Sastre, I. Dornic, and H. Chat\'e,
\texttt{cond-mat/0308178}.

\bibitem{gl02} C. Godr\`eche and J.M. Luck, J. Phys. Cond. Matt. {\bf
14}, 1589 (2002).

\bibitem{salerno2} F. Corberi, E. Lippiello, and M. Zannetti,
\texttt{cond-mat/0307542}.

\bibitem{pleimling} M. Pleimling, \texttt{cond-mat/0309652}.

\bibitem{malte2} M. Henkel, M. Paessens, and M. Pleimling,
\texttt{cond-mat/0310761}.

\bibitem{salerno3} F. Corberi, C. Castellano, E. Lippiello, and
M. Zannetti, \texttt{cond-mat/0311046}.

\bibitem{kawa} K. Kawasaki, Phys. Rev. {\bf 145}, 224 (1966).

\bibitem{maj} S.N. Majumdar, D.A. Huse, and B.D. Lubachevsky,
Phys. Rev. Lett. {\bf 73}, 182 (1994).

\bibitem{alexander} F.J. Alexander, D.A. Huse, and S.A. Janowsky,
Phys. Rev. B {\bf 50}, 663 (1994).

\bibitem{satya} S.N. Majumdar and D.A. Huse,
Phys. Rev. E {\bf 52}, 270 (1995).

\bibitem{yeung} C. Yeung, M. Rao, and R.C. Desai, Phys. Rev. E {\bf
53}, 3073 (1996).

\bibitem{salerno1} F. Corberi, E. Lippiello, and M. Zannetti,
Phys. Rev. E {\bf 65}, 066114 (2002).

\bibitem{chatelain} C. Chatelain, J. Phys. A {\bf 36}, 10739 (2003).

\bibitem{fede} F.~Ricci-Tersenghi, Phys. Rev. E {\bf 68}, 065104(R)
(2003).

\bibitem{cks} S.J. Cornell, K. Kaski, and R.B. Stinchcombe,
Phys. Rev. B {\bf 44}, 12263 (1991).

\bibitem{smedt} G. De Smedt, C. Godr\`eche, and J.M.~Luck,
Eur. Phys. J. B {\bf 32}, 215 (2003).

\bibitem{cordery} R. Cordery, S. Sarker, and J. Tobochnik,
Phys. Rev. B {\bf 24}, 5402 (1981).

\bibitem{glKawa} C. Godr\`eche and J.M.~Luck, J. Phys. A \textbf{36},
9973 (2003).

\bibitem{husefisher} D.S. Fisher and D.A. Huse, Phys. Rev. B {\bf 38},
373 (1988).

\bibitem{amar} J.G. Amar, F.E. Sullivan, and R.D. Mountain, 
Phys. Rev. B {\bf 37}, 196 (1988).

\end{thebibliography}
\end{document}